\documentclass[10pt,letterpaper,conference]{IEEEtran}

\usepackage{amsmath,amssymb,amsfonts}
\usepackage{gensymb}
\usepackage{array}
\usepackage{algorithmic}
\usepackage{graphicx}
\usepackage{textcomp}
\usepackage{xcolor}
\usepackage{xspace}
\usepackage{enumitem}
\usepackage{url}
\usepackage{breakurl}
\usepackage{caption}
\usepackage{subcaption}
\usepackage{multirow}
\usepackage[normalem]{ulem}
\usepackage[numbers,sort&compress]{natbib}
\usepackage[linesnumbered,ruled,lined,boxed]{algorithm2e}
\graphicspath{{figures/}}
\SetKwInput{KwInput}{Input}                
\SetKwInput{KwOutput}{Output}              

\newcolumntype{L}[1]{>{\raggedright\let\newline\\\arraybackslash\hspace{0pt}}m{#1}}
\newcolumntype{C}[1]{>{\centering\let\newline\\\arraybackslash\hspace{0pt}}m{#1}}
\newcolumntype{R}[1]{>{\raggedleft\let\newline\\\arraybackslash\hspace{0pt}}m{#1}}

\usepackage{graphicx}
\usepackage{booktabs}
\usepackage{multirow}
\usepackage{tabularx}
\usepackage{subcaption}
\usepackage{amsmath, amssymb}
\usepackage{xcolor}
\usepackage{caption}

\usepackage{amsthm}
\usepackage{amsmath}
\usepackage{amsfonts}
\usepackage{mathtools}
\usepackage{epstopdf}
\usepackage{array}
\usepackage{csquotes}

\usepackage{longtable}
\begin{document}


\title{Wi-Fi Self-Coexistence in the 6~GHz Band: An ns-3 Evaluation of LPI and SP Usage}

\author{
\IEEEauthorblockN{
Hossein Nasiri\IEEEauthorrefmark{1},
Seda Dogan-Tusha\IEEEauthorrefmark{1},
Francis A. Gatsi\IEEEauthorrefmark{2},
and Monisha Ghosh\IEEEauthorrefmark{1}\\
\IEEEauthorblockA{
\IEEEauthorrefmark{1}Department of Electrical and Electronics Engineering, University of Notre Dame, South Bend, IN, USA\\}
\IEEEauthorrefmark{2}Department of Computer Science and Engineering, University of Notre Dame, South Bend, IN, USA\\}

\IEEEauthorblockA{
Email: \{hnasiri2, stusha, fgatsi, mghosh3\}@nd.edu}
}


\maketitle

\begin{abstract}
The U.S. has adopted four power regimes for operation in the shared unlicensed 6~GHz band---standard power (SP), low-power indoor (LPI), geofenced variable power (GVP), and very low power (VLP)---with maximum permitted EIRP levels of 36~dBm, 30~dBm, 24~dBm, and 14~dBm, respectively. Although these regimes are primarily intended to protect incumbent services, their heterogeneous transmit power levels also introduce additional coexistence challenges within 6~GHz Wi-Fi networks. In this paper, we develop an ns-3 Wi-Fi~6E/802.11ax coexistence testbed to study coexistence under heterogeneous power regimes and to provide a reproducible simulation methodology. To the best of our knowledge, prior work has not specifically examined self-coexistence issues within 6~GHz Wi-Fi networks. We evaluate two coexistence scenarios: one in which both the LPI AP and the SP AP are indoors, and another in which the LPI AP is indoors while the SP AP is outdoors. Results are compared against an indoor LPI--LPI baseline when applicable. Our findings show that: (i) the presence of an indoor SP AP can significantly degrade the goodput of an LPI AP; (ii) channel bandwidth is a key factor in determining the extent of SP-to-LPI impact, with the degradation being most severe at 20~MHz and partially alleviated at 160~MHz; (iii) physical blockage between outdoor SP and LPI APs improves fairness; and (iv) BSS coloring does not necessarily improve fairness in mixed-regime deployments. The simulation framework can be extended to study coexistence between Wi-Fi and cellular systems, as recently proposed by Ofcom in the U.K.
\end{abstract}



\section{Introduction}\label{introduction}

The 6~GHz band (5.925--7.125~GHz) hosts incumbent services such as fixed microwave links and satellite systems, which require strict interference protection \cite{FCC1}. To meet the increasing demand for unlicensed spectrum while safeguarding these incumbents, the U.S.\ Federal Communications Commission (FCC) authorized unlicensed operations in this band in 2020, subject to a set of regulatory constraints \cite{FCC2}.

\begin{figure}[t]
 \centering
 \includegraphics[width=\linewidth]{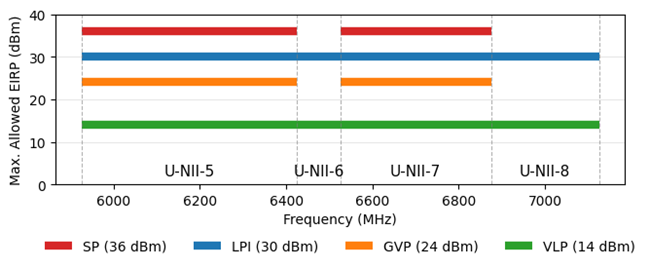}
 \caption{Maximum allowed EIRP levels in the 6~GHz band.}
 \label{fig:6ghz_freq_chart}
\end{figure}

The FCC has established multiple classes of unlicensed devices operating in the 6~GHz band, each with distinct power limits and operational requirements \cite{FCC2, FCC3}. Low-power indoor (LPI) devices are allowed to operate throughout the entire band without coordination, provided they are used indoors and adhere to a maximum equivalent isotropically radiated power (EIRP) limit of 30~dBm and a power spectral density (PSD) limit of 5~dBm/MHz. Standard power (SP) devices may transmit at higher power levels, with a maximum EIRP of 36~dBm and PSD of 23~dBm/MHz, but must employ an Automated Frequency Coordination (AFC) system to prevent harmful interference with incumbent users. Very-low-power (VLP) devices, intended for short-range and portable applications, operate with a maximum EIRP of 14~dBm and a PSD limit of $-1$~dBm/MHz. In January 2026, the FCC has adopted a new class of higher-power geofenced variable power (GVP) devices capable of outdoor operation with up to 11 dBm/MHz PSD and 24 dBm EIRP \cite{fcc6ghz_gvp_2026_v1}. Figure~\ref{fig:6ghz_freq_chart} illustrates the EIRP limits for these device classes across the Unlicensed National Information Infrastructure (U-NII) bands, while Table~\ref{tab:regulation_6e} summarizes the maximum EIRP and PSD limits across various channel bandwidths. Client devices must maintain transmit power at least 6~dB below the access point's (AP) limits. Notably, for 20~MHz channels, SP APs can transmit at power levels up to 18~dB higher than those of LPI APs; however, this power gap narrows with increasing bandwidth because of the fixed EIRP cap applied to SP APs. 

The coexistence of heterogeneous power regimes in the 6 GHz band leads to diverse transmission ranges and interference characteristics. The adoption of GVP devices highlights the increasing need to understand coexistence behavior among unlicensed devices with varying power constraints, especially in mixed indoor–outdoor deployments.  Furthermore, in January 2026, Ofcom proposed sharing between Wi-Fi and cellular systems in the upper 6~GHz band in the U.K.~\cite{Ofcom6ghz_2026}, where the two coexisting networks will be operating at widely different power levels, necessitating further study of such scenarios.


\begin{table}[t]
\centering
\renewcommand{\arraystretch}{1.1}
\caption{Maximum EIRP for 6~GHz SP, LPI, and VLP APs with respect to channel bandwidth.}
\label{tab:regulation_6e}
\begin{tabularx}{\columnwidth}{c *{5}{>{\centering\arraybackslash}X}}
\toprule
\multicolumn{6}{c}{\textbf{Max Tx Power (dBm)}} \\
\midrule
\textbf{AP Type} & 20~MHz & 40~MHz & 80~MHz & 160~MHz & 320~MHz \\
\midrule
SP (PSD: 23~dBm/MHz) & 36 & 36 & 36 & 36 & 36 \\
GVP (PSD: 11~dBm/MHz) & 24 & 24 & 24 & 24 & 24 \\
LPI (PSD: 5~dBm/MHz) & 18 & 21 & 24 & 27 & 30 \\
VLP (PSD: $-5$~dBm/MHz) & 12 & 14 & 14 & 14 & 14 \\
\bottomrule
\end{tabularx}
\end{table}

Prior coexistence studies have largely focused on symmetric deployments or on Wi-Fi coexistence with cellular technologies such as LTE-LAA and NR-U, where differences in energy detection thresholds and channel access mechanisms are known to create unfairness and inefficiencies~\cite{RochmanEDThreshold, Merhnoush2018WiFiLAA, keshtiarast2025_coexistence, sathya2021measurement}.
In~\cite{RochmanEDThreshold, Bellalta2019SpatialReuse}, the authors highlight that power asymmetry fundamentally alters the effectiveness of spatial reuse mechanisms. When neighboring transmitters operate at significantly different power levels, carrier sensing becomes asymmetric. 
In such scenarios, higher-power transmitters may fail to sense lower-power activity, leading to increased interference and inefficient channel usage, while lower-power devices continue to defer conservatively.


The coexistence of heterogeneous power regimes in the unlicensed 6~GHz band presents notable challenges for fair and efficient Wi-Fi self-coexistence. Current 6~GHz deployments predominantly consist of SP and LPI APs. LPI APs were initially concentrated in university campus environments but have since expanded broadly across enterprise deployments~\cite{dogan2023evaluating,dogan2023indoor,10856858}. Following the FCC's authorization of AFC in the 6~GHz band, SP AP deployments have increased, particularly in outdoor scenarios, owing to their higher transmit power and larger coverage areas.
Our prior work~\cite{dogantusha2025} characterizes the interactions between SP and LPI devices and the potential for interference to incumbents through extensive measurement campaigns conducted at the University of Notre Dame. The results show that dense outdoor SP deployments can negatively impact indoor LPI performance. Despite the growing interest in 6~GHz Wi-Fi networks, system-level studies that explicitly examine coexistence between SP and LPI APs remain limited in the literature.

\subsection{Scope and Contributions}\label{sec:contrib}
Given the discussion above, this paper presents a system-level ns-3 evaluation of LPI and SP usage in the 6~GHz band and makes the following contributions:

\begin{itemize}
 \item We construct an ns-3 Wi-Fi 6E/802.11ax coexistence testbed for heterogeneous power regimes and provide a reproducible simulation methodology, including parameterized sweeps over channel bandwidth, modulation and coding scheme (MCS), offered load, and overlapping basic service set packet detect (OBSS-PD) thresholds. Using this framework, we evaluate two coexistence scenarios: one in which both the LPI AP and the SP AP are indoors (referred to as the indoor LPI--SP scenario), and another in which the LPI AP is indoors while the SP AP is outdoors (referred to as the indoor--outdoor LPI--SP scenario). Results are compared against an indoor LPI--LPI baseline.

 
 \item We show that indoor LPI--SP coexistence is unfair relative to the indoor LPI--LPI baseline: LPI APs experience significant goodput degradation due to a transmit power imbalance, with the impact being most severe at 20~MHz, partially alleviated at 160~MHz, and further intensified under higher-MCS, high-rate operation.

 \item We demonstrate that indoor--outdoor LPI--SP coexistence improves fairness relative to indoor scenarios. This is primarily due to the isolation provided by walls, which reduces the airtime gap between the SP and LPI APs.
 

 \item We provide an analysis showing that when LPI and SP APs are within 100~m, airtime usage is balanced due to deferral, with increased disparity beyond 100~m; the airtime gap is also smaller when the SP AP is outdoors.

 \item We study whether Basic Service Set (BSS) coloring/OBSS-PD can mitigate SP dominance, or whether aggressive thresholds can harm performance.
 
\end{itemize}

\section{Essential Background: Why SP--LPI Coexistence Is Different}\label{sec:background}

Unlike classical ``two symmetric Wi-Fi networks'' coexistence, LPI--SP coexistence combines (i) different EIRP limits, (ii) different coverage footprints, and (iii) potentially different sensing outcomes at APs and STAs due to location and receiver sensitivity. In Carrier Sense Multiple Access with Collision Avoidance (CSMA/CA) networks, fairness is tightly coupled to who senses whom and for how long. In mixed-power deployments, SP devices can (i) be sensed earlier and from greater distances, (ii) capture channel access opportunities by sustaining higher SINR links (i.e., higher successful transmission probability), and (iii) indirectly bias association and throughput by reducing the effective channel availability for LPI networks.

We emphasize that unfairness here is evaluated against an indoor LPI--LPI reference, not merely between AP0 and AP1 within the same run. That is, we interpret unfairness as \emph{``how much LPI performance degrades when the neighboring AP is an SP instead of an LPI''} under an otherwise identical topology and offered load.

\section{System Model and Methodology}\label{sec:system_model}

\subsection{Simulation Overview}\label{sec:sim_overview}
\begin{figure}[t]
 \centering
 \includegraphics[width=0.7\linewidth]{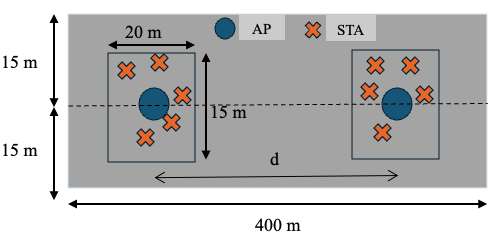}
 \caption{Deployment for LPI and indoor SP APs: The gray shaded area denotes the indoor area.} 
 \label{fig:indoor-setup}
\end{figure}

We simulate a two-AP deployment using ns-3 Wi-Fi 802.11ax (Wi-Fi 6E) with a configurable channel bandwidth $B\in\{20, 80, 160\}$~MHz. Data transmissions use an MCS $\in\{\mathrm{HeMcs0}, \mathrm{HeMcs5}, \mathrm{HeMcs9}\}$, while control frames are transmitted using $\mathrm{HeMcs0}$. 
The scenario contains two APs (AP0 and AP1). Each AP serves five STAs placed uniformly at random within a local $20 \times 15$~m region centered on the AP. This ensures an identical offered load per AP and keeps STAs close to their serving APs. Both APs and STAs are equipped with two transmit and two receive antennas. Traffic consists of downlink UDP flows from the APs to their associated STAs. We consider two scenarios: 

\paragraph{\textbf{Indoor co-channel coexistence}}
The indoor scenario shown in Fig.~\ref{fig:indoor-setup} consists of two neighboring APs placed in a rectangular region of size $400~\mathrm{m} \times 30~\mathrm{m}$
with the long dimension enabling evaluation at large separations. The two APs are located along the $x$-axis at $(200 - \frac{d}{2}, 15)$ and $(200 + \frac{d}{2}, 15)$, where $d$ is the distance between the two APs, swept from 40 to 360~m in steps of 20~m.

Two indoor cases are evaluated: (i) LPI--LPI, where both APs operate under LPI power constraints and serve as the baseline, and (ii) LPI--SP, where one AP operates as an LPI and the other as an SP.

\begin{figure}[t]
 \centering
 \includegraphics[width=0.7\linewidth]{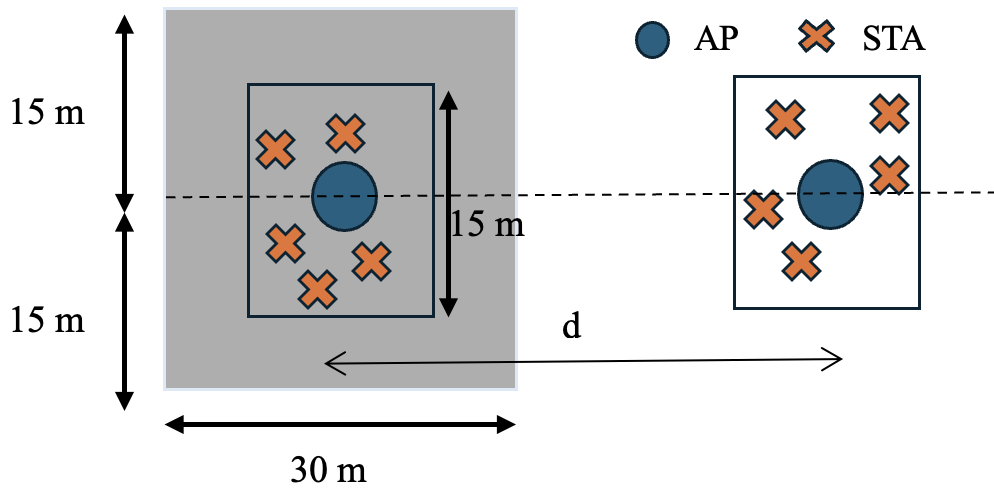}
 \caption{Deployment for LPI and outdoor SP APs: The gray shaded area denotes the indoor area, while the white area denotes the outdoor area.}
 \label{fig:outdoor-setup}
\end{figure}

\paragraph{\textbf{Indoor--outdoor co-channel coexistence}} The indoor--outdoor LPI--SP scenario models a mixed deployment in which an SP AP is located outdoors and an LPI AP operates indoors, as shown in Fig.~\ref{fig:outdoor-setup}. The indoor environment is represented by a $30 \times 30$~m building, with the LPI AP at its center while the SP AP is placed outside the building. The separation between the SP and LPI APs is swept from 40 to 360~m in steps of 20~m. It is noteworthy that an outdoor LPI--LPI baseline is not considered, since LPI operation is restricted to indoor environments. 

Furthermore, all APs operate with identical PHY and MAC configurations. Wireless propagation is modeled using the ns-3 \texttt{YansWifiChannel} with a \texttt{HybridBuildingsPropagationLossModel}, enabling combined distance-based pathloss and building-related attenuation. Indoor environments are configured using an office building model, while outdoor-to-indoor scenarios incorporate penetration loss through the same hybrid propagation framework. Transmit power levels for LPI and SP APs are set according to bandwidth-dependent regulatory limits, which are summarized in Table~\ref{tab:regulation_6e}.


\begin{table*}[t]
\caption{ns-3 simulation configuration for LPI--LPI and LPI--SP coexistence scenarios.}
\label{tab:ns3_params}
\centering
\small
\renewcommand{\arraystretch}{1.0}
\begin{tabular}{p{3.3cm} p{5.5cm} p{6.8cm}}
\toprule
\textbf{Category} & \textbf{Parameter} & \textbf{Value} \\
\midrule
Topology
& Region / layout
& $400 \times 30$~m, 2 APs, 5 STAs/AP (random scatter inside a $20 \times 15$~m area centered at the AP) \\
& AP separation
& 40--360~m (20~m step)\\
& Simulation time / runs
& 10~s, 10 independent seeds \\

\midrule
PHY / Rate
& Standard / band
& IEEE 802.11ax (Wi-Fi 6E) \\
& BW 
& 20, 80, 160~MHz \\
& MCS
& Data: HeMcs0, HeMcs5, HeMcs9; Control: HeMcs0 \\
& Antennas
& AP: $2\times2$ MIMO; STA: $2\times2$ MIMO \\
& Traffic direction
& Downlink UDP \\

\midrule
Power regimes
& TX power configuration
& LPI and SP AP/STA powers set per FCC limits (Table~\ref{tab:regulation_6e}) \\
& Scenario selection
& LPI--LPI vs.\ LPI--SP \\

\midrule
MAC / Channel access
& RTS/CTS
& Disabled for all frames (\texttt{rtsThresh}=0) \\
& OBSS-PD
& Enabled/disabled; BSS colors = $\{1, 2\}$ when active \\
& Other MAC parameters
& Enhanced Distributed Channel Access (EDCA), A-MPDU at ns-3 defaults \\

\midrule
Traffic
& Packet size / rate
& 1200~bytes, CBR, $\Delta t=100~\mu$s \\

\midrule
Outputs
& Metrics
& Per-STA goodput, latency, airtime ratio, Rx packets, lost packets \\

\bottomrule
\end{tabular}
\end{table*}

\subsection{Traffic Model and Offered Load}\label{sec:traffic_model}
We generate downlink UDP traffic from each AP to its associated stations using a constant bit rate (CBR) model. All results in this paper are based exclusively on CBR traffic to ensure controlled and repeatable offered-load conditions across scenarios.

Each AP transmits fixed-size packets of length $L=1200$~bytes with a constant inter-packet interval of $\Delta t=100~\mu$s. This corresponds to a per-flow offered load of
\begin{equation}
R_{\mathrm{CBR}} = \frac{8L}{\Delta t} = \frac{8 \times 1200}{100 \times 10^{-6}} = 96~\mathrm{Mbps}.
\end{equation}

With five stations associated to each AP, the aggregate offered downlink load per AP is approximately $480$~Mbps. We find that this load is sufficient to drive the network into a contention-limited regime for most bandwidth and modulation configurations considered, allowing coexistence effects between LPI and SP deployments to be clearly observed.

\subsection{AP Association}\label{sec:assoc}
In this study, association is predetermined. All STAs are explicitly assigned to their respective APs at initialization and remain associated with those APs for the entire simulation duration. This configuration is equivalent to a different-SSID deployment and ensures that each AP serves an identical number of STAs under the same offered traffic load.



\subsection{OBSS-PD and BSS Coloring}\label{sec:cca_obss}
We evaluate spatial reuse mechanisms defined in 802.11ax:
\begin{itemize}
 \item \textbf{BSS Coloring:} The two APs are assigned different BSS colors.
 \item \textbf{OBSS-PD:} 
 Under IEEE 802.11ax, the OBSS-PD mechanism allows a receiver to ignore inter-BSS frames below a configured OBSS-PD threshold for physical carrier sensing, enabling increased spatial reuse by permitting more aggressive channel access.
\end{itemize}
We therefore parameterize OBSS-PD threshold values ($-72$ and $-62$~dBm) and compare them against the case in which OBSS-PD is disabled. Importantly, OBSS-PD changes the trade-off between spatial reuse and interference/collision risk; aggressive thresholds can increase concurrent transmissions but may also increase packet loss and reduce throughput.

\subsection{RTS/CTS and A-MPDU Configuration}\label{sec:rts}
Request--to--send/clear--to--send (RTS/CTS) is disabled in all simulations to reflect common default operation and to avoid additional control overhead. Since RTS/CTS can interact with OBSS-PD and BSS coloring by modifying channel access and Network Allocation Vector (NAV) behavior, keeping it disabled allows us to isolate coexistence effects driven by power asymmetry and carrier sensing rather than control-plane mechanisms.

Aggregate MAC Protocol Data Unit (A-MPDU) is enabled and configured with the maximum allowed aggregation size, allowing up to 4,194,304 bytes per aggregated transmission as specified by the IEEE 802.11ax standard. Enabling A-MPDU reflects typical high-throughput Wi-Fi configurations and allows each successful channel access to deliver multiple data frames within a single transmission opportunity (TXOP).

\subsection{Metrics}\label{sec:metrics}

The main metrics used to evaluate performance are goodput, Jain's fairness across APs goodput, latency, and airtime ratio as a proxy for channel occupancy. We compute:
\begin{itemize}
 \item \textbf{Per-AP aggregate goodput:} 
 \begin{equation}
 G_{\mathrm{AP}}(d) = \sum_{i \in \mathcal{S}_{\mathrm{AP}}} G_{i}(d),
 \end{equation}
 where $d$ is the distance between the APs and $G_{i}$ denotes goodput in Mbps for STA $i$, defined as the rate of successfully received MAC-layer payload bits, excluding protocol overhead and retransmissions. 

 \item \textbf{Jain's fairness index} across the two APs:
 \begin{equation}
 J(d) = \frac{\left(G_{\mathrm{AP0}}(d) + G_{\mathrm{AP1}}(d)\right)^2}
 {2\left(G_{\mathrm{AP0}}(d)^2 + G_{\mathrm{AP1}}(d)^2\right)}.
 \end{equation}
 This metric captures the balance of useful data delivery between the two APs and reflects fairness in channel access under coexistence.

 \item \textbf{Airtime ratio:} Defined as the fraction of the simulation time during which an AP actively occupies the channel for transmission,
 \begin{equation}
 A_{\mathrm{AP}} = \frac{T_{\mathrm{tx}}}{T_{\mathrm{sim}}},
 \end{equation}
 where $T_{\mathrm{tx}}$ denotes the cumulative PHY transmission time of the AP and $T_{\mathrm{sim}}$ is the total simulation duration. A larger airtime ratio indicates longer channel access and a greater opportunity to deliver useful data.

\item \textbf{Latency:} The average end-to-end per-packet delay experienced by each STA.

\end{itemize}


\subsection{Statistical Treatment}\label{sec:stats}
Each configuration is run for 10 random seeds. For each distance point $d$, we average the results over all runs. 
 To ensure that the observed performance differences are attributable to power-regime asymmetry and channel-access dynamics rather than to mismatched PHY or load parameters, we verify that:
 
 \begin{itemize}
 \item per-AP goodput is computed as the sum of goodputs of the STAs associated with that AP,
 \item all LPI--LPI and LPI--SP comparisons are performed under identical bandwidth, MCS, and traffic configurations.
\end{itemize}
 


\subsection{Simulation Parameters}\label{sec:sim_params_detailed}

Table~\ref{tab:ns3_params} reports the detailed ns-3 configuration. This table is intentionally verbose to support reproducibility and reviewer scrutiny.

\section{Measurement-Based Observations from Real-World Deployment}\label{sec:measurement_context}

In \cite{dogantusha2025}, our recent work presents an extensive measurement campaign for Wi-Fi~6E deployment at the University of Notre Dame Stadium (NDS) in South Bend, IN, USA. At NDS, outdoor SP APs are deployed to provide wide-area coverage, while nearby buildings host indoor LPI APs operating in the unlicensed 6~GHz band. This deployment highlights the practical and growing prevalence of LPI--SP APs coexistence, demonstrating that such configurations are not exceptional but representative of emerging deployment environments.

\begin{figure}
 \centering \includegraphics[width=0.8\linewidth]{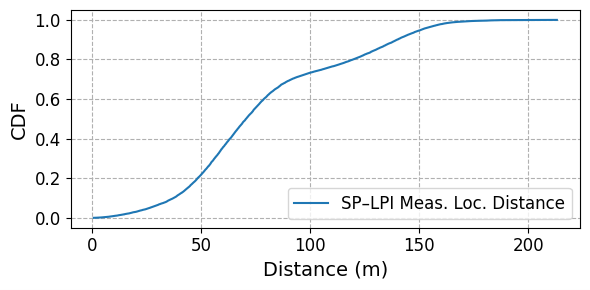}
 \caption{CDF of pairwise distances between all (SP, LPI) pairs on the same channel. }
 \label{fig:meas_sp_lpi_meas_distance}
\end{figure}

As part of a measurement campaign, we collected signal-level data from both SP and LPI networks. Figures 14(b) and 15(b) in \cite{dogantusha2025} reveal a high density of SP BSSIDs operating on the same channel as LPI in the U-NII-5 and U-NII-7 bands, particularly at indoor locations near windows. To better understand the practical proximity between SP and LPI APs and to inform our ns-3 modeling, we examine the distribution of distances between co-channel SP and LPI APs. Specifically, we analyze measurements collected from one SP AP and one LPI AP operating on the same channel, compute pairwise Euclidean distances between their measurement locations, and present the resulting distribution as a cumulative distribution function (CDF) in Fig.~\ref{fig:meas_sp_lpi_meas_distance}. Although this approach does not yield the exact physical separation between the APs, it provides a practical indication of their relative proximity. The estimated separation distances between SP and LPI span a wide range from near co-location to approximately 200~m. These observations motivate the distance sweep used in our ns-3 evaluation. To ensure that both near-field and far-field coexistence effects are captured, our simulation study explores separation distances up to 
400~m. This allows us to examine not only the regimes directly observed in the measurements but also extended-distance behaviors.


\section{Results and Discussion: ns-3 Evaluation}\label{sec:results}

In this section, we evaluate LPI AP coexistence with SP AP indoor and outdoor using ns-3. We examine goodput, Jain’s fairness index, and average airtime to quantify LPI performance in the presence of SP and benchmark it against the LPI--LPI baseline. Table~\ref{tab:scenario_matrix} summarizes the scenarios evaluated in this work.


\subsection{Indoor SP AP impact on LPI AP}
\label{sec:short_distance}

This section first examines indoor co-channel coexistence and quantifies LPI AP performance degradation in the presence of a SP AP. All indoor results are interpreted relative to a homogeneous LPI--LPI baseline, which represents the expected behavior when both APs operate under identical power constraints. We analyze how LPI goodput evolves as a function of inter-AP distance, channel bandwidth, and modulation.

\begin{table}[t]
\caption{Scenario matrix for ns-3 simulation evaluation.}
\label{tab:scenario_matrix}
\centering
\small
\renewcommand{\arraystretch}{1.05}
\begin{tabular}{p{2.8cm} p{4.6cm}}
\toprule
\textbf{Scenario} & \textbf{Configuration details} \\
\midrule
Baseline LPI--LPI 
& Both APs operate under LPI regime \\

Indoor LPI--SP 
& One LPI AP and one SP AP indoors \\

Indoor--outdoor LPI--SP 
& One LPI AP indoors and one SP AP outdoors \\



Traffic 
& DL UDP CBR, 1200~bytes packets, 100~$\mu$s interval \\

STAs 
& 5 STAs per AP (10 total) \\

MAC features 
& A-MPDU enabled; RTS/CTS disabled \\

OBSS-PD 
& Disabled and enabled with threshold sweep \\
\bottomrule
\end{tabular}
\end{table}


\begin{figure*}[t]
 \centering
 \begin{subfigure}{0.32\linewidth}
 \centering
 \includegraphics[width=\linewidth]{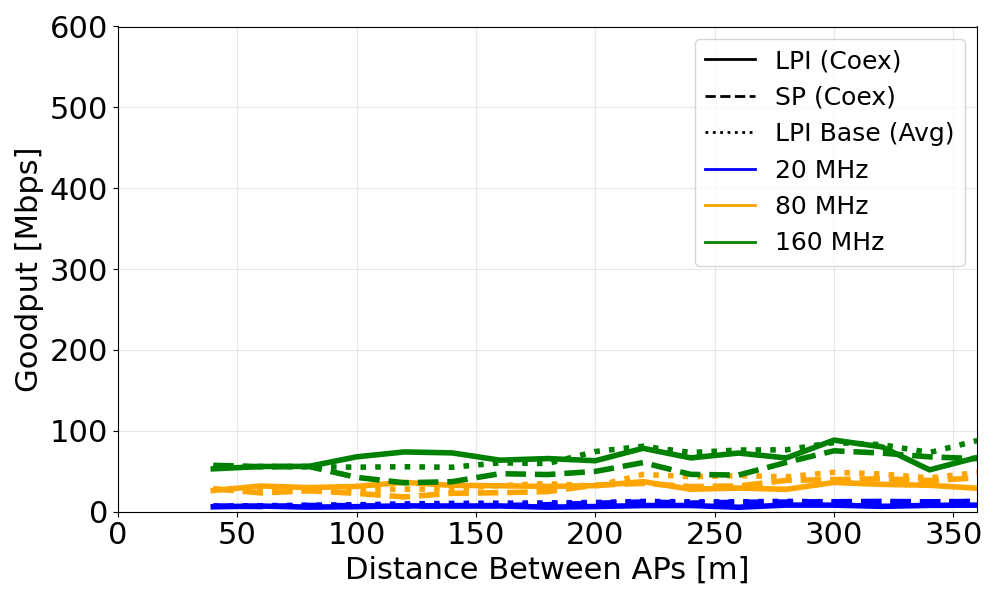}
 \caption{MCS~0}
 \label{fig:indoor_MCS0_goodput}
 \end{subfigure}
 \begin{subfigure}{0.32\linewidth} 
 \centering
 \includegraphics[width=\linewidth]{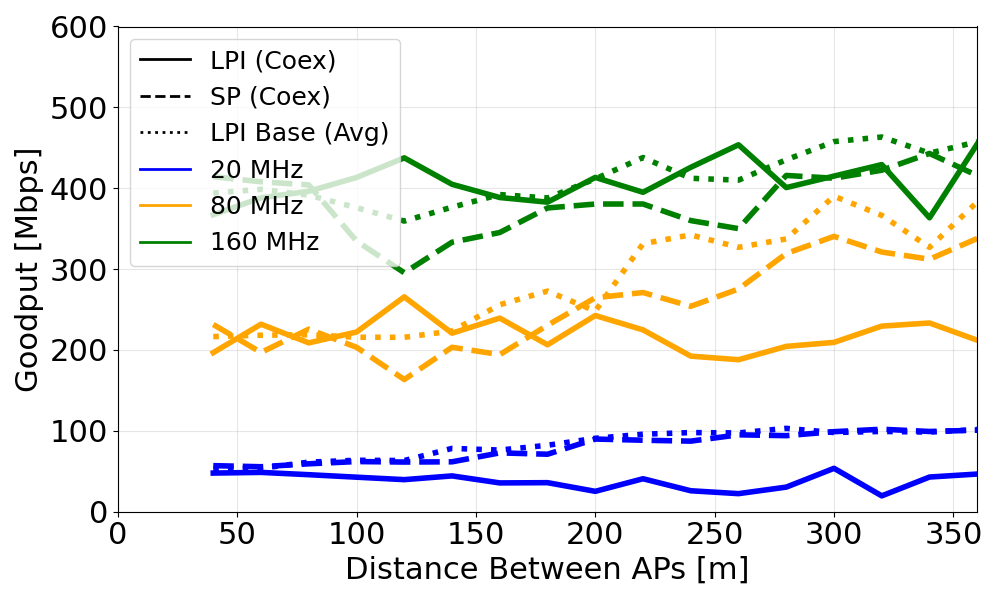}
 \caption{MCS~5}
 \label{fig:indoor_MCS5_goodput}
 \end{subfigure}
 \begin{subfigure}{0.32\linewidth} 
 \centering
 \includegraphics[width=\linewidth]{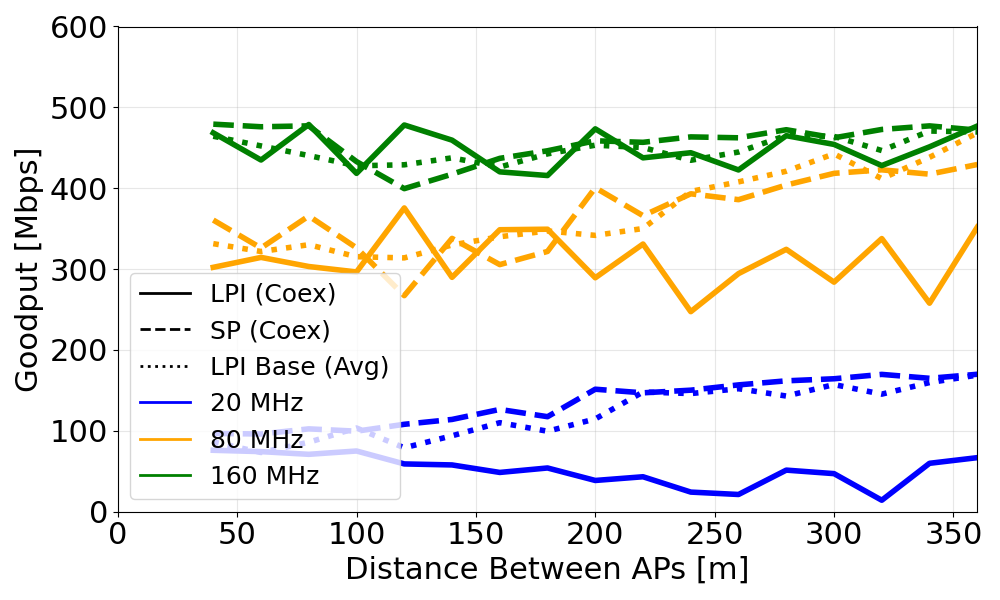}
 \caption{MCS~9}
 \label{fig:indoor_MCS9_goodput}
 \end{subfigure}
 \caption{Goodput versus distance for MCS0, MCS5, and MCS9, comparing SP AP throughput, LPI AP throughput under indoor LPI--SP coexistence, and the LPI--LPI baseline average.} 
 \label{fig:indoor_goodput}
\end{figure*}

\begin{figure*}[t]
 \centering
 \begin{subfigure}{0.32\linewidth}
 \includegraphics[width=\linewidth]{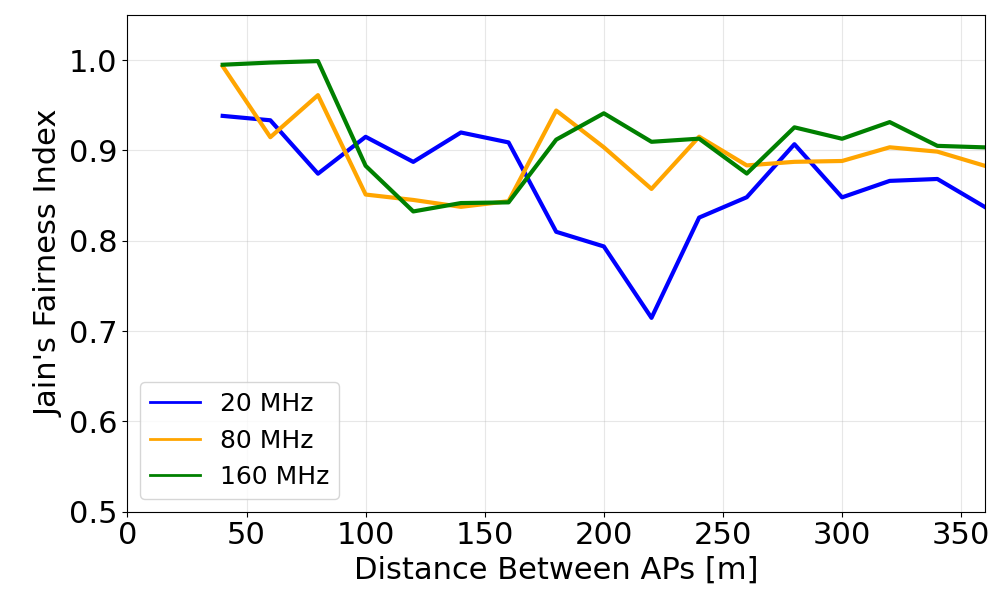}
 \caption{MCS~0}
 \label{fig:mcs0_fairness_on}
 \end{subfigure}\hfill
 \begin{subfigure}{0.32\linewidth}
 \includegraphics[width=\linewidth]{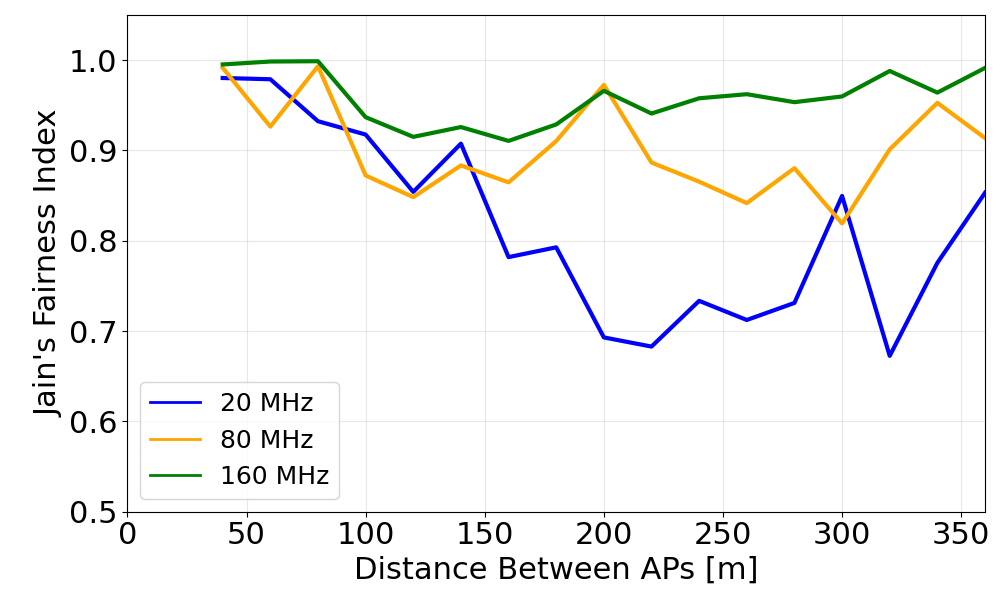}
 \caption{MCS~5}
 \label{fig:mcs5_fairness_on}
 \end{subfigure}\hfill
 \begin{subfigure}{0.32\linewidth}
 \includegraphics[width=\linewidth]{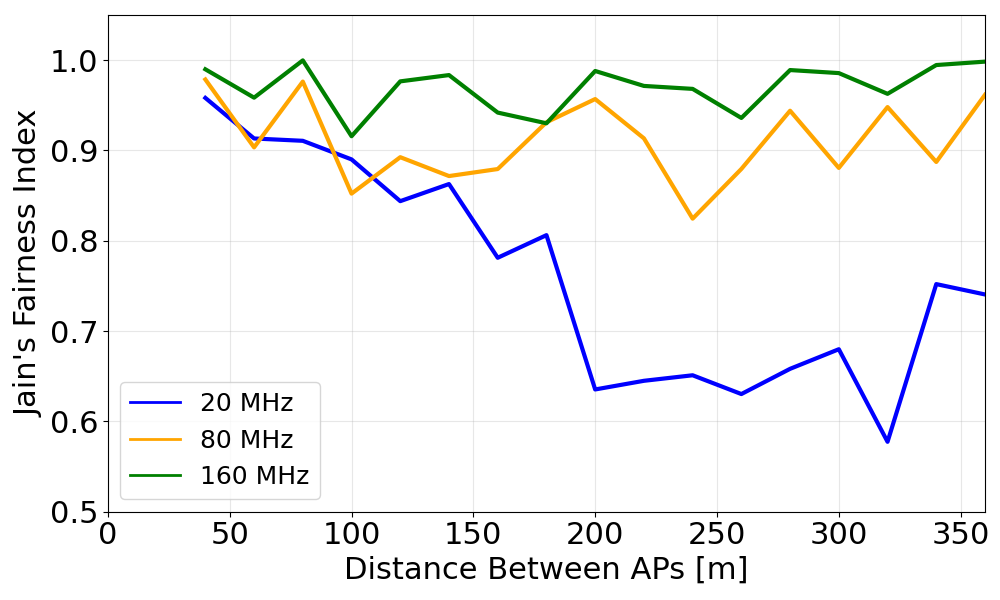}
 \caption{MCS~9}
 \label{fig:mcs9_fairness_on}
 \end{subfigure}\hfill
 \caption{Jain's fairness index between the LPI and SP APs versus distance for indoor LPI--SP coexistence at MCS0, MCS5, and MCS9.} 
 \label{fig:mcs_fairness}
\end{figure*}

\begin{figure}[t]
 \centering
 \begin{subfigure}{0.45\linewidth} 
 \centering
 \includegraphics[width=\linewidth]{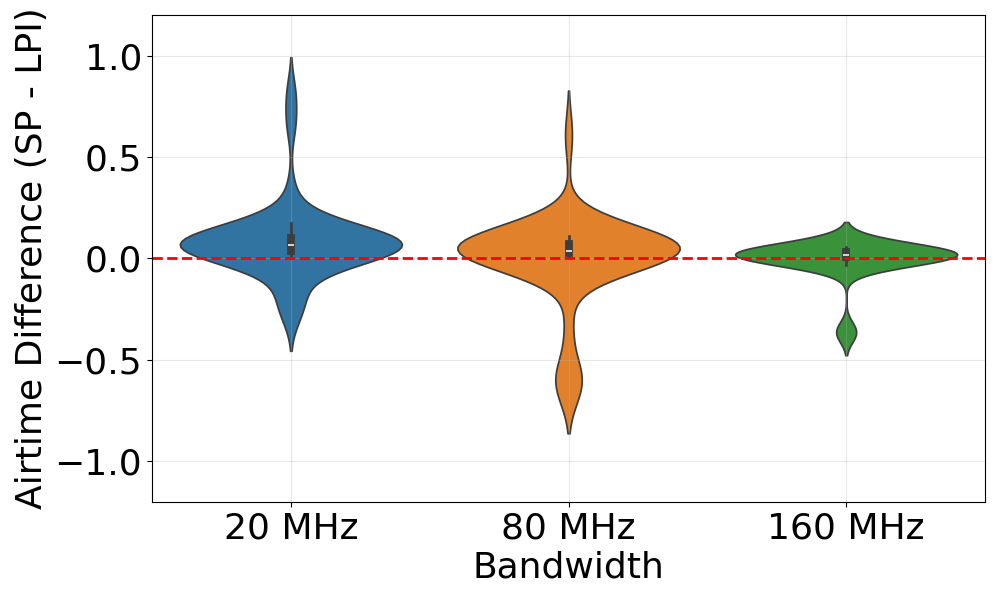}
 \caption{\( d \leq 100 \, \text{m} \)}
 \label{fig:airtime_indoor_dbelow100}
 \end{subfigure}
 \begin{subfigure}{0.45\linewidth} 
 \centering
 \includegraphics[width=\linewidth]{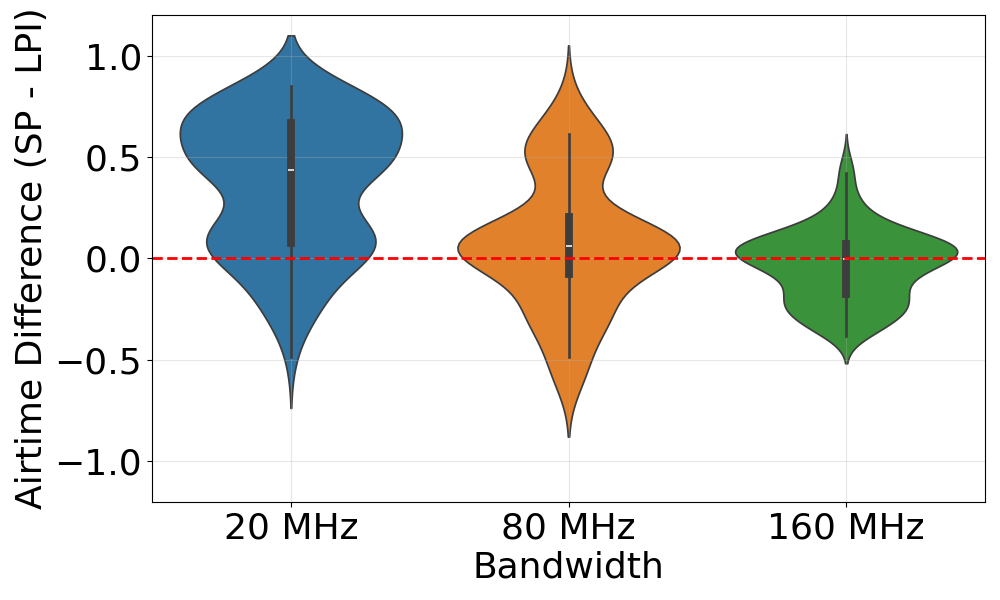}
 \caption{\( d \geq 100 \, \text{m} \) }
 \label{fig:airtime_indoor_dabove100}
 \end{subfigure}
 \caption{Distribution of the airtime-ratio difference for indoor LPI--SP coexistence at MCS~5, shown separately for \( d \leq 100 \, \text{m} \) and \( d \geq 100 \, \text{m} \).} 
 \label{fig:air_time_diff}
\end{figure}

\begin{figure*}[t]
 \centering
 \begin{subfigure}{0.32\linewidth}
 \includegraphics[width=\linewidth]{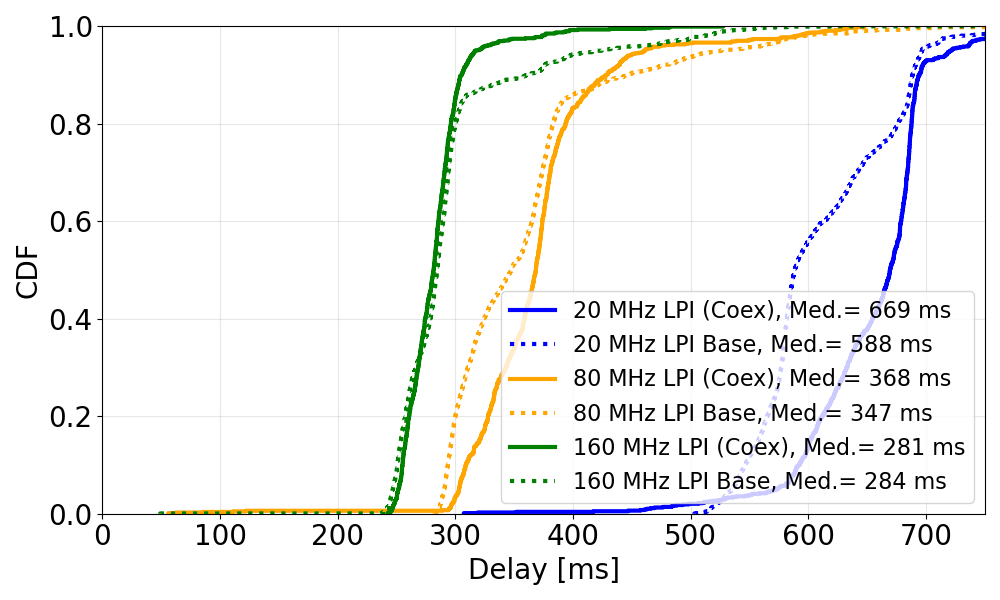}
 \caption{MCS~0}
 \label{fig:mcs0_latency}
 \end{subfigure}\hfill
 \begin{subfigure}{0.32\linewidth}
 \includegraphics[width=\linewidth]{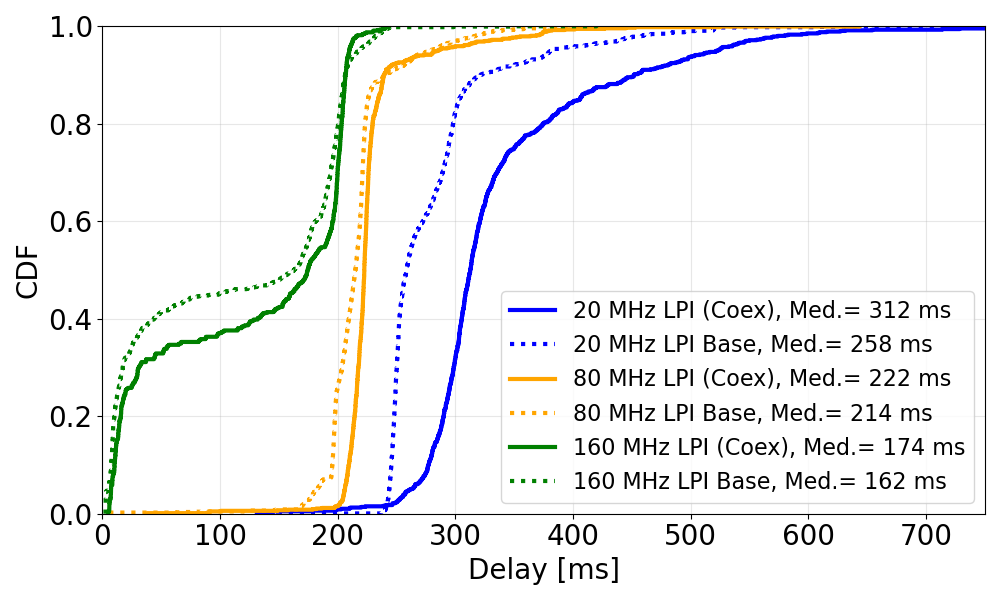}
 \caption{MCS~5}
 \label{fig:mcs5_latency}
 \end{subfigure}\hfill
 \begin{subfigure}{0.32\linewidth}
 \includegraphics[width=\linewidth]{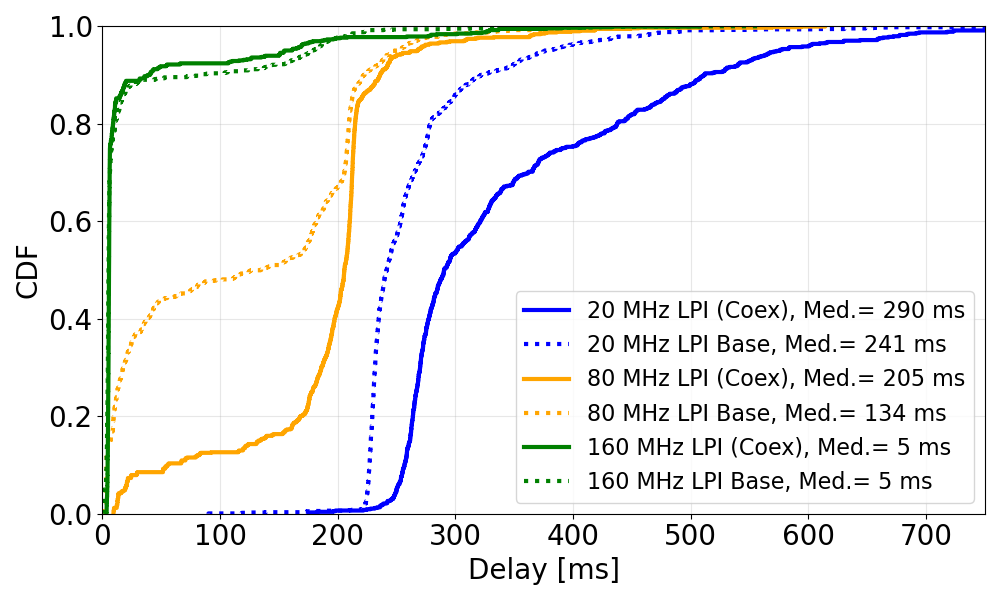}
 \caption{MCS~9}
 \label{fig:mcs9_latency}
 \end{subfigure}

 \caption{CDF of average latency (ms) for LPI-associated users under indoor LPI--SP coexistence and the LPI--LPI baseline.}
 \label{fig:mcs_latency}
\end{figure*}

\begin{figure}[t]
 \centering
 \begin{subfigure}{0.48\linewidth}
 \centering
 \includegraphics[width=\linewidth]{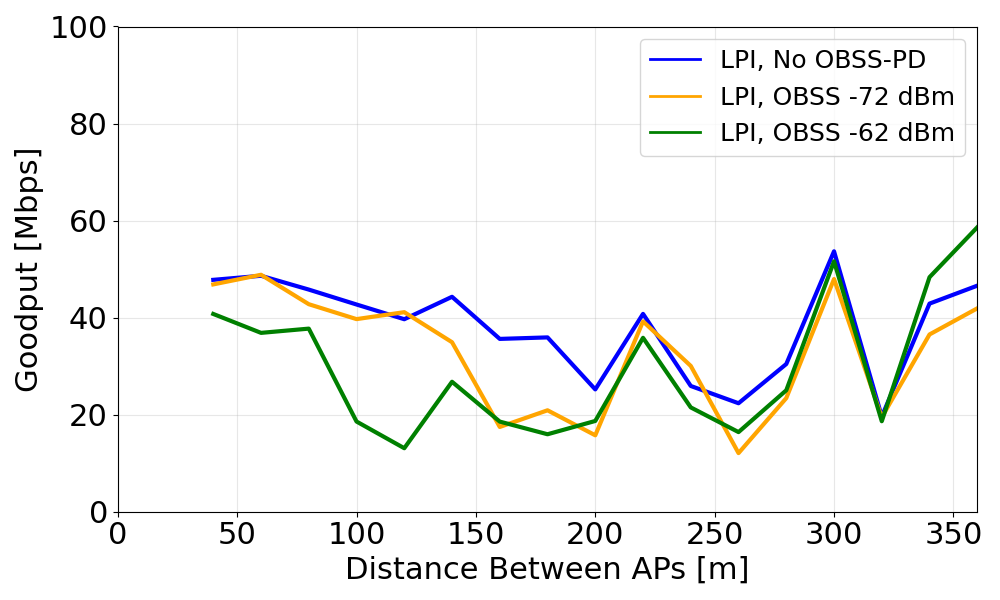}
 \caption{MCS5 }
 \label{fig:BSS_indoor_MCS5}
 \end{subfigure}\hfill
 \begin{subfigure}{0.48\linewidth}
 \centering
 \includegraphics[width=\linewidth]{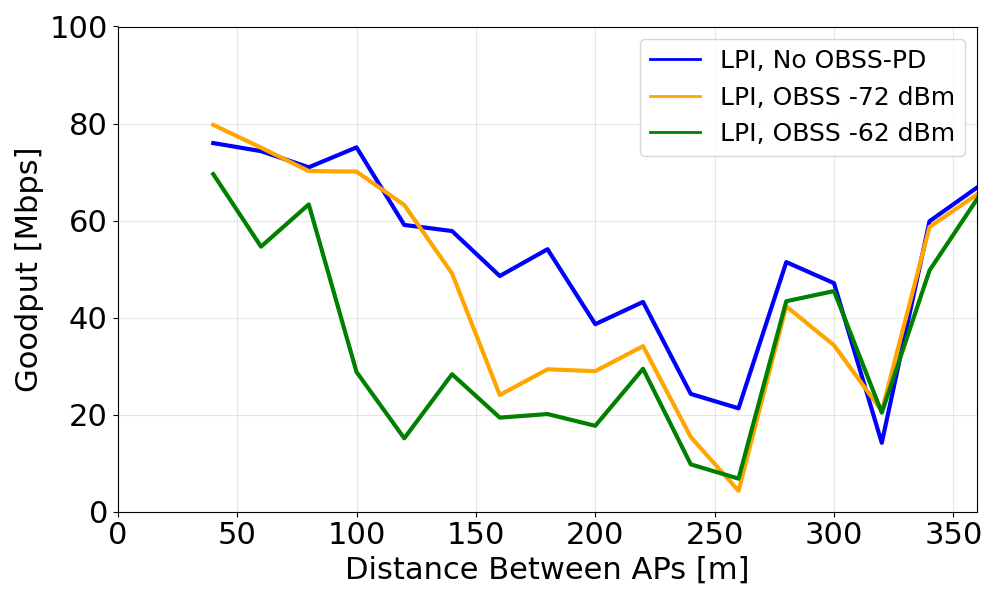}
 \caption{MCS9 }
 \label{fig:BSS_indoor_MCS9}
 \end{subfigure}
 \caption{LPI goodput versus distance for indoor LPI--SP coexistence under OBSS-PD at 20~MHz for MCS~5 and MCS~9.} 
 \label{fig:BSS_indoor_20MHz}
\end{figure}

\subsubsection{Goodput Performance}
Fig.~\ref{fig:indoor_goodput} plots goodput versus inter-AP distance for 20~MHz, 80~MHz, and 160~MHz channels at MCS~0, MCS~5, and MCS~9, comparing the SP AP, the LPI AP under mixed-regime operation, and the LPI--LPI baseline average. MCS~0 provides a robust PHY with limited sensitivity to channel variations and contention, MCS~5 represents a mid-rate operating point, and MCS~9 captures high-rate transmission that is more sensitive to coexistence-driven contention and deferrals. 

From Fig.~\ref{fig:indoor_goodput}, two key observations emerge.
First, LPI AP performance degradation in the presence of an SP AP becomes more pronounced as the MCS increases. Specifically, MCS~0 (Fig.~\ref{fig:indoor_MCS0_goodput}) exhibits the smallest degradation in goodput, while MCS~9 (Fig.~\ref{fig:indoor_MCS9_goodput}) shows the largest degradation. This trend reflects the increased sensitivity of higher-rate MCS configurations to interference and channel access disruptions.

Second, channel bandwidth has a significant impact on coexistence behavior. At 20~MHz, the transmit power difference between the SP AP and the LPI AP is 18~dB, leading to a clear dominance of the SP AP in channel access. In contrast, at 160~MHz, where the transmit power difference is reduced to approximately 9~dB due to regulatory EIRP limits, the goodput gap between the SP AP and the LPI AP is noticeably smaller. This indicates that wider bandwidths partially mitigate LPI AP degradation by reducing the effective power asymmetry between coexisting APs.

Beyond MCS and bandwidth, the distance between APs ($d$) is a critical determinant of coexistence quality. Fig.~\ref{fig:indoor_MCS0_goodput} illustrates that at MCS~0, goodput remains low and weakly dependent on distance. In contrast, Fig.~\ref{fig:indoor_MCS5_goodput} demonstrates that at MCS~5, performance becomes highly sensitive to $d$. Specifically, for $d \leq 100$~m, degradation is minimal across all bandwidths because the APs remain within mutual carrier-sensing range, ensuring balanced channel access similar to the LPI--LPI baseline. However, as separation increases beyond 100~m, the significant power deficit in 20 and 80~MHz channels causes the LPI AP to defer to the SP AP while the SP AP fails to detect LPI transmissions, leading to severe goodput loss. In particular, 160~MHz channel mitigates this asymmetry due to the reduced power difference, maintaining fairer coexistence. 



\subsubsection{Jain's Fairness Index}

To quantify fairness between the SP AP and the LPI AP, we report Jain's fairness index as a function of distance for different MCS values and channel bandwidths, as shown in Fig.~\ref{fig:mcs_fairness}. 

As shown in Fig.~\ref{fig:mcs0_fairness_on}, when MCS~0 is used, Jain’s fairness for the 20~MHz channel remains relatively high and rarely drops below 0.8 (as observed between 200 and 240 m) in the distance range evaluated. This indicates relatively balanced throughput between the two networks and reflects more equitable channel access. In contrast, for higher-rate configurations, fairness degrades more noticeably. Specifically, for MCS~5 and MCS~9, shown in Fig.~\ref{fig:mcs5_fairness_on} and Fig.~\ref{fig:mcs9_fairness_on}, respectively, the fairness index for the 20~MHz channel drops below 0.8 just beyond approximately 150~m. The results also show that wider channel bandwidths improve channel access equity, yielding a more even throughput distribution and suggesting that wider bandwidths improve fairness in these scenarios.

\subsubsection{Airtime Ratio Comparison}


%

Fig.~\ref{fig:air_time_diff} illustrates the probability density function (PDF) of the airtime ratio difference,
\begin{equation}
\Delta A = A_{\mathrm{SP}} - A_{\mathrm{LPI}},
\end{equation}
where \( A_{\mathrm{SP}} \) and \( A_{\mathrm{LPI}} \) denote the airtime ratios of the SP AP and the LPI AP, respectively. We plot the distributions separately for short separations (\( d \leq 100 \)~m, Fig.~\ref{fig:airtime_indoor_dbelow100}) and larger separations (\( d \geq 100 \)~m, Fig.~\ref{fig:airtime_indoor_dabove100}) for MCS~5. In mixed LPI--SP coexistence, the dominant AP is expected to exhibit a higher airtime ratio than its counterpart. At $d\geq100$~m, the distribution for the 20~MHz channel becomes strongly skewed toward positive values, indicating persistent dominance of the SP AP in channel access. This behavior is consistent with the pronounced goodput degradation observed for LPI at narrow bandwidths and large separations. The increasing goodput gap and reduced fairness at $d\geq 100$ m observed in Figs. \ref{fig:indoor_goodput} and \ref{fig:mcs_fairness} can be directly attributed to this airtime imbalance between the SP AP and the LPI AP.

In contrast, for wider bandwidths---particularly 160~MHz---the airtime ratio difference varies much less with distance. This reduced imbalance reflects the smaller transmit power difference imposed by regulatory EIRP limits at wider bandwidths and explains the milder LPI performance degradation observed in earlier results. 

\subsubsection{Latency Performance}

Beyond goodput, delay is a vital metric for assessing network performance. Low latency is a prerequisite for a smooth user experience in widely used applications like video streaming, VoIP, and augmented and virtual reality (AR/VR).

As shown in Fig.~\ref{fig:mcs_latency}, we analyze the average latency of STAs per AP, excluding inactive STAs with goodput below 0.1~Mbps. The results show that LPI-associated STAs in the indoor LPI--SP coexistence setup experience higher latency than in the LPI--LPI baseline, especially for MCS~5 and MCS~9 at 20~MHz. However, increasing the system bandwidth narrows this performance gap, which is consistent with the goodput trends discussed above.

\subsubsection{BSS Coloring and OBSS-PD Effects}\label{sec:bss}

The power imbalance between SP and LPI APs, which results in unequal airtime allocation, raises the question of whether spatial-reuse mechanisms can improve coexistence fairness. In particular, we investigate whether asymmetric BSS coloring can compensate for this imbalance by making the LPI AP more aggressive in channel access. To this end, we assign distinct BSS colors to the LPI and SP networks and increase the OBSS-PD threshold only for the LPI AP, while keeping the SP AP sensing configuration unchanged at $-82$~dBm. This design allows the LPI AP to transmit under more relaxed sensing conditions, potentially increasing its channel-access opportunities.

We focus on the most challenging coexistence configurations—MCS~5 and MCS~9 with a 20~MHz channel—where fairness degradation is most pronounced. Fig.~\ref{fig:BSS_indoor_20MHz} compares LPI goodput in the LPI--SP coexistence scenario under three configurations: OBSS disabled, OBSS-PD set to $-72$~dBm, and OBSS-PD set to $-62$~dBm at the LPI AP. The results show that increasing the OBSS-PD threshold does not improve LPI goodput. Instead, more aggressive thresholds lead to further goodput degradation.

These observations indicate that, in the presence of strong power asymmetry, relaxing carrier sensing at the LPI AP primarily increases exposure to SP AP transmissions rather than enabling effective spatial reuse. Consequently, asymmetric BSS coloring is insufficient to restore fairness or improve LPI performance in this regime.


\subsection{Outdoor SP Impact on Indoor LPI Performance}\label{sec:outdoor}

We now consider an indoor--outdoor deployment scenario in which, as shown in Fig.~\ref{fig:outdoor-setup}, the SP access point is located outside the building, while the LPI access point and its associated stations operate indoors. Unlike the indoor scenario, an LPI--LPI baseline is not feasible in this case, as LPI devices are restricted to indoor operation by regulation. Consequently, results in this section are interpreted relative to the indoor LPI reference behavior established in the previous sections.


\begin{figure}[t]
 \centering
 \begin{subfigure}{0.45\linewidth} 
 \centering
 \includegraphics[width=\linewidth]{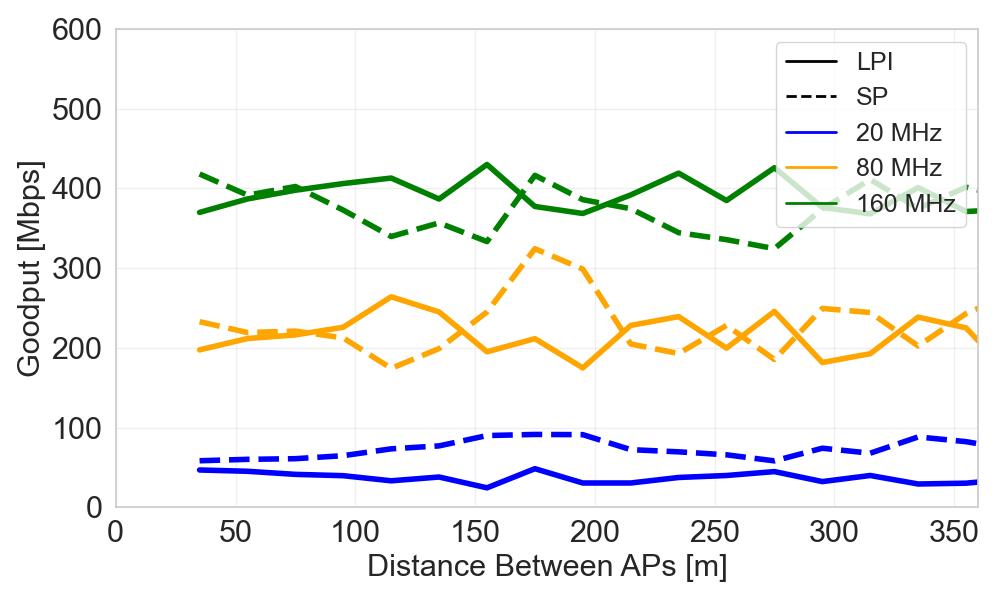}
 \caption{MCS~5}
 \label{fig:out_thpt_MCS5}
 \end{subfigure}
 \begin{subfigure}{0.45\linewidth} 
 \centering
 \includegraphics[width=\linewidth]{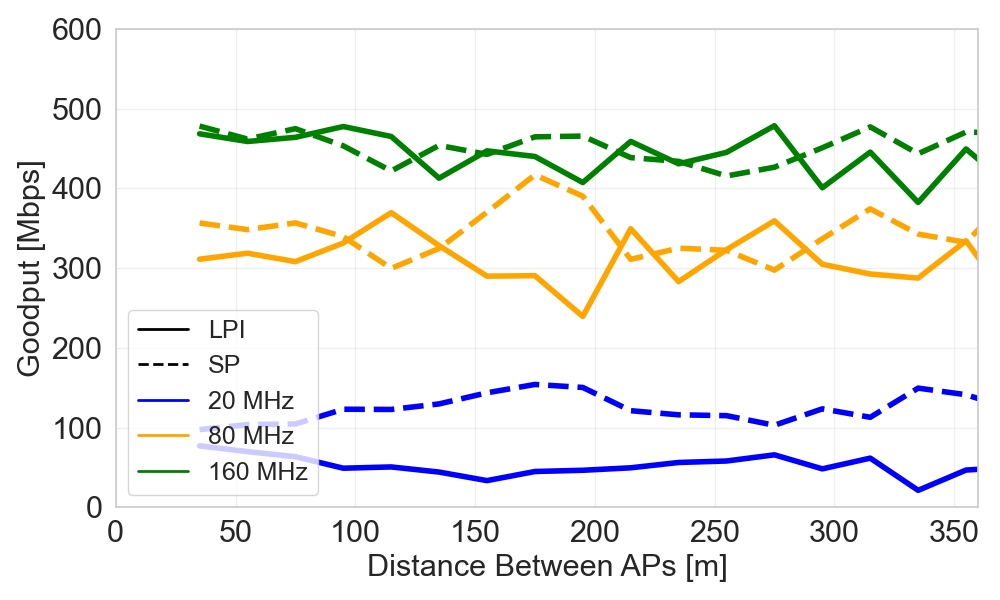}
 \caption {MCS9}
 \label{fig:out_thpt_MCS9}
 \end{subfigure}
 \caption{Goodput versus distance for MCS~5 and MCS~9, comparing SP throughput, LPI throughput under indoor--outdoor LPI--SP coexistence, and the LPI--LPI baseline average.} 
 \label{fig:out_thpt}
\end{figure}

\subsubsection{Goodput Performance}
Fig.~\ref{fig:out_thpt} shows LPI AP and SP AP goodput versus distance for the indoor--outdoor LPI--SP scenario across different channel bandwidths. Although the building introduces penetration loss, the outdoor SP AP continues to affect the performance of indoor LPI AP over a wide range of distances. LPI AP's goodput remains particularly suppressed for the 20~MHz bandwidth.



As in the indoor case, bandwidth plays a critical role. At narrower bandwidths, LPI goodput degradation is more pronounced due to asymmetric carrier sensing, where indoor LPI devices continue to defer to SP transmissions while the outdoor SP no longer consistently senses LPI activity. Increasing the bandwidth reduces the severity of this effect. Since the SP transmit power is capped by regulation and does not scale proportionally with bandwidth, the relative power asymmetry between SP and LPI decreases, allowing indoor LPI goodput to recover more effectively at wider bandwidths.
\subsubsection{Jain's Fairness Index}
\begin{figure}[t]
 \centering
 \begin{subfigure}{0.48\linewidth}
 \centering
 \includegraphics[width=\linewidth]{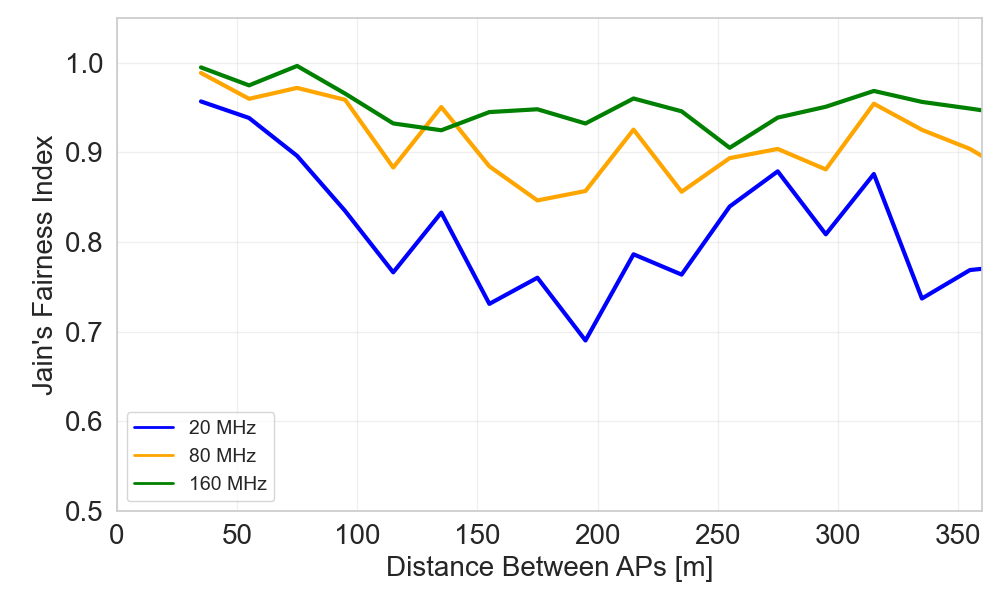}
 \caption{MCS~5}
 \label{fig:outdoor_MCS5_fairness}
 \end{subfigure}\hfill
 \begin{subfigure}{0.48\linewidth}
 \centering
 \includegraphics[width=\linewidth]{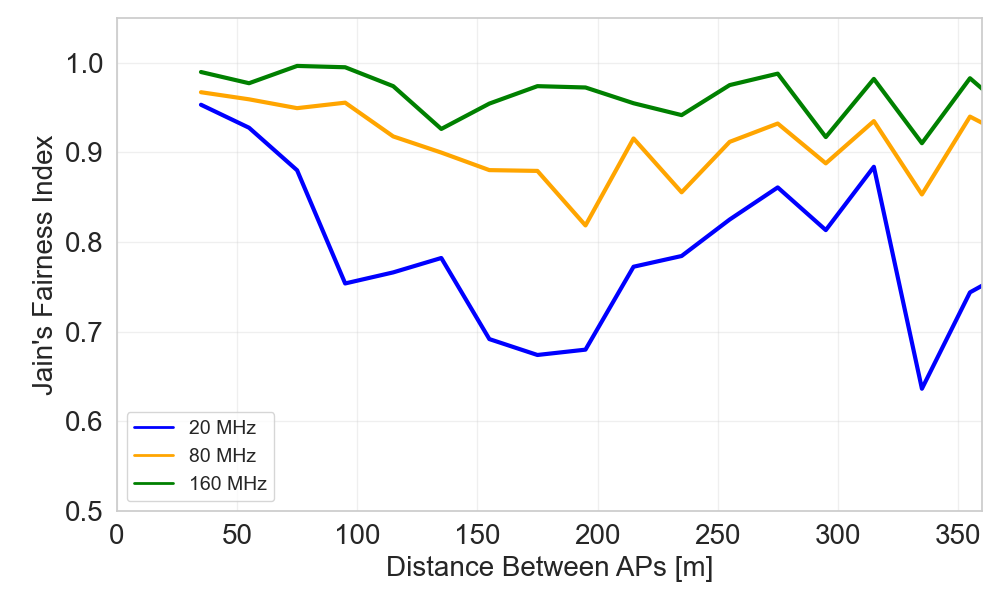}
 \caption{MCS~9}
 \label{fig:outdoor_MCS9_fairness}
 \end{subfigure}
 \caption{Jain's fairness index versus distance for indoor--outdoor LPI--SP coexistence.} 
 \label{fig:outdoor_fairness}
\end{figure}

As shown in Fig.~\ref{fig:outdoor_fairness}, the results confirm that channel bandwidth plays a crucial role in fairness. For both MCS~5 and MCS~9, wider channel bandwidths yield higher values of Jain's fairness index. This indicates a more even throughput distribution for wider bandwidths. The index for 20 MHz also increases beyond approximately 200~m, reflecting the weakened impact of the SP AP on the indoor LPI network as the distance between the two APs increases. 

\subsubsection{Airtime Ratio Comparison}
\begin{figure}[t]
 \centering
 \begin{subfigure}{0.45\linewidth} 
 \centering
 \includegraphics[width=\linewidth]{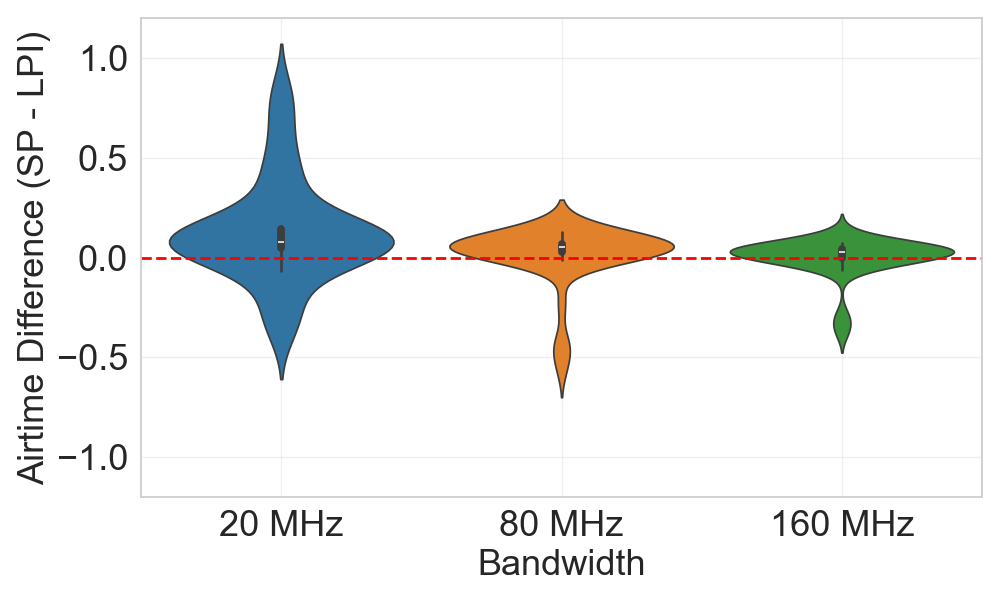}
 \caption{\( d \leq 100 \, \text{m} \)}
 \label{fig:airtime_outdoor_dbelow100}
 \end{subfigure}
 \begin{subfigure}{0.45\linewidth} 
 \centering
 \includegraphics[width=\linewidth]{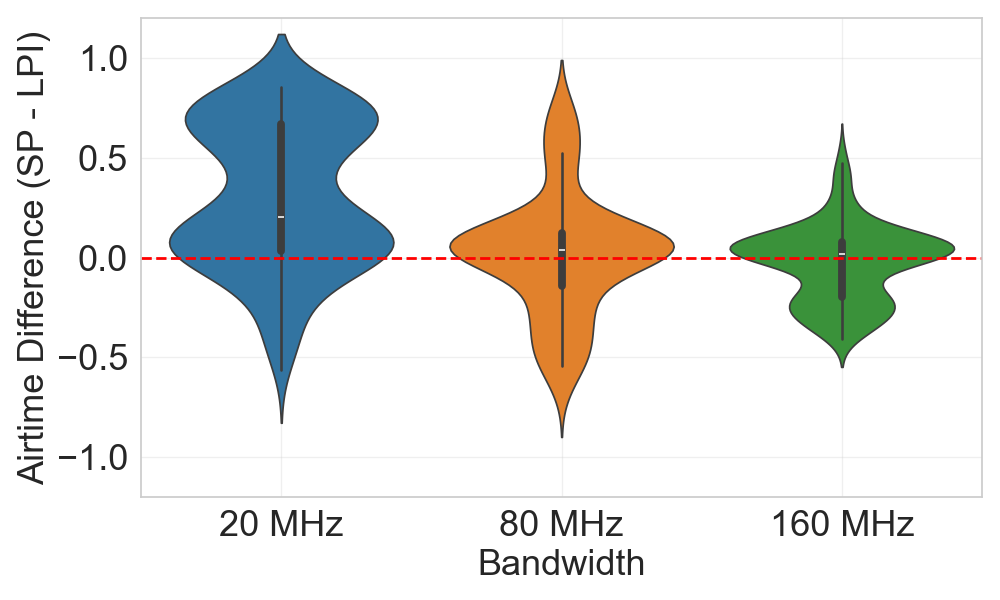}
 \caption{\( d \geq 100 \, \text{m} \) }
 \label{fig:airtime_outdoor_dabove100}
 \end{subfigure}
 \caption{Distribution of the airtime-ratio difference for indoor--outdoor LPI--SP coexistence at MCS~5, shown separately for \( d \leq 100 \, \text{m} \) and \( d \geq 100 \, \text{m} \).} 
 \label{fig:air_time_diff-out}
\end{figure}

Similar to the indoor scenario, asymmetric (one-directional) carrier sensing at larger separations leads to a growing imbalance in channel access between the SP AP and the LPI AP. As illustrated in Fig.~\ref{fig:airtime_outdoor_dbelow100} and Fig.~\ref{fig:airtime_outdoor_dabove100}, the distribution of the airtime ratio difference between the SP AP and the LPI AP is increasingly skewed toward SP dominance at distances above 100~m, compared to the short-range regime. This confirms that sensing asymmetry caused by transmit power imbalance leads to uneven channel occupancy at larger separations. 

\section{Conclusions and Future Work}\label{sec:conclusion}
We presented a detailed ns-3 system-level evaluation of coexistence between LPI and SP Wi-Fi 6E APs in the 6~GHz band. Using an LPI--LPI baseline, we showed that replacing one LPI AP with an SP AP can substantially degrade LPI performance, and that the severity depends on bandwidth, PHY rate, offered load, and association dynamics. We further demonstrated that an outdoor SP AP can still affect indoor LPI networks in a wall/blockage scenario. Furthermore, we observed that, although BSS coloring/OBSS-PD is often cited as a mitigation technique, its impact in these heterogeneous scenarios is limited and can exacerbate fairness issues rather than resolve them. Future work will extend the analysis to denser multi-AP topologies and examine coexistence scenarios involving VLP devices alongside LPI and SP deployments. In addition, we will consider a broader range of deployment scenarios, such as office and residential environments, various building types and layouts (including multi-floor structures), and mixed-use settings to better capture realistic coexistence conditions in the 6~GHz band.

\section*{\centering{Acknowledgments}}
This research was funded in part by NSF Grant \#CNS-2229387.

\bibliographystyle{IEEEtran}
\bibliography{main}

@misc{Ofcom6ghz_2026,
  author = {{Ofcom}},
  title  = {{Ofcom takes strides towards mobile and {Wi-Fi} sharing same airwaves in boost for {UK} economy}},
  month  = {Jan.},
  year   = {2026}
}

@INPROCEEDINGS{dogantusha2025,
  author={Dogan-Tusha, Seda and others},
  booktitle={2025 IEEE International Symposium on Dynamic Spectrum Access Networks (DySPAN)}, 
  title={{Evaluation of Indoor/Outdoor Sharing in the Unlicensed 6 GHz Band}}, 
  year={2025},
  volume={},
  number={},
  pages={1-9},
  doi={10.1109/DySPAN64764.2025.11115954}}

@ARTICLE{10856858,
  author={Doğan-Tusha, Seda and others},
  journal={IEEE Communications Magazine}, 
  title={{Spectrum Sharing Characterization Using Smartphones: Exploring 6 GHz Sharing Through Large-Scale Wi-Fi 6E Measurements}}, 
  year={2025},
  volume={63},
  number={2},
  pages={70-76},
  doi={10.1109/MCOM.001.2400325}}

@misc{fcc6ghz_gvp_2026_v1,
  author       = {{Federal Communications Commission}},
  title        = {{Announcement on expansion of unlicensed 6\,GHz operations and new GVP devices}},
  month        = {Jan.},
  year         = {2026},
  url          = {https://docs.fcc.gov/public/attachments/DOC-417577A1.pdf}
}

@inproceedings{keshtiarast2025_coexistence,
  title        = {{Coexistence analysis of Wi-Fi 6E and 5G NR-U in the 6 {GHz} band}},
  author       = {Keshtiarast, Navid and Petrova, Marina},
  booktitle    = {Proceedings of the 2025 International Conference on ns-3 (ICNS3 ’25)},
  year         = {2025},
  pages        = {38--45},
isbn = {9798400715174},
publisher = {Association for Computing Machinery},
}

@article{sathya2021measurement,
  title        = {{Measurement‑based coexistence studies of LAA \& Wi‑Fi deployments in Chicago}},
  author       = {Vanlin Sathya and Muhammad Iqbal Rochman and Monisha Ghosh},
  journal      = {IEEE Wireless Communications},
  volume       = {28},
  number       = {1},
  pages        = {136--143},
  year         = {2021},
  doi          = {10.1109/MWC.001.2000205},
}

@article{RochmanEDThreshold,
 author={Iqbal Rochman, Muhammad and others},
  booktitle={Wireless Telecommunications Symposium (WTS)}, 
  title={{Impact of changing energy detection thresholds on fair coexistence of Wi-Fi and LTE in the unlicensed spectrum}}, 
  year={2017},
  volume={},
  number={},
  pages={1-9},
  doi={10.1109/WTS.2017.7943527}
}

@article{Merhnoush2018WiFiLAA,
  author  = {Merhnoush, M. and others},
  journal={IEEE/ACM Transactions on Networking}, 
  title={{Analytical modeling of Wi-Fi and LTE-LAA coexistence: Throughput and impact of energy detection threshold}}, 
  year={2018},
  volume={26},
  number={4},
  pages={1990-2003},
}

@misc{FCC1,
author = {{FCC News Release}},
title = {{FCC opens 6 GHz band to Wi-Fi and other unlicensed uses}},
month = {Apr.},
note={{Accessed: Sep. 2025}},
howpublished={Retrieved from \url{https://www.fcc.gov/document/fcc-opens-6-ghz-band-wi-fi-and-other-unlicensed-uses}}
}

@misc{FCC2,
author = {{FCC, GN Docket No. 17-183}},
title = {{FCC adopts new rules for the 6 GHz band, unleashing 1200 megahertz of spectrum for unlicensed use}},
year = {2020},
month = {Apr.},
note={{Accessed: Sep. 2025}},
howpublished={Retrieved from \url{https://docs.fcc.gov/public/attachments/DOC-363945A1.pdf}}
}

@misc{FCC3,
author = {{FCC Fact Sheet}},
title = {{Exploring flexible use in mid-band spectrum between 3.7 GHz and 24 GHz 
}},
year = {2017},
month = {Jul.},
note= {{Accessed: Sep. 2025}},
howpublished={Retrieved from \url{https://docs.fcc.gov/public/attachments/FCC-20-51A1.pdf}}
}

@inproceedings{dogan2023evaluating,
  title={{Evaluating the interference potential in 6 {GHz}: An extensive measurement campaign of a dense indoor {Wi-Fi} {6E} network}},
  author={Dogan-Tusha, Seda and others},
  booktitle={Proceedings of the 17th ACM Workshop on Wireless Network Testbeds, Experimental evaluation \& Characterization},
  pages={56--63},
  year={2023}
}

@inproceedings{dogan2023indoor,
  title={{Indoor and outdoor Measurement Campaign for Unlicensed 6 GHz Operation with Wi-Fi 6E}},
  author={Dogan-Tusha, Seda and others},
  booktitle={International Symposium on Wireless Personal Multimedia Communications (WPMC)},
  year={2023},
  pages={1--6},
  organization={IEEE}
}

@article{Bellalta2019SpatialReuse,
  author  = {Bellalta, Boris},
  title   = {{IEEE 802.11ax: High-Efficiency WLANs}},
  journal = {IEEE Wireless Communications},
  volume  = {26},
  number  = {1},
  pages   = {140--147},
  year    = {2019},
  month   = {February},
  doi     = {10.1109/MWC.2019.1800155}
}

\end{document}